\documentclass{ws-procs9x6}
\usepackage{graphicx}

\begin{document}

\title{Properties of neutron stars}

\author{M.~H.~van Kerkwijk}

\address{Department of Astronomy \& Astrophysics, \\
60 Saint George Street, \\ 
Toronto, Ontario, M5S 3H8, Canada\\ 
E-mail: mhvk@astro.utoronto.ca}

\maketitle

\abstracts{I review attempts made to determine the properties of
neutron stars.  I focus on constraints on the maximum mass that a
neutron star can have, and on attempts to measure neutron-star radii.
So far, there appears to be only one neutron star for which there is
strong evidence that its mass is above the canonical $1.4\,M_\odot$,
viz., Vela~X-1, for which a mass close to $1.9\,M_\odot$ is found.
Prospects for progress appear brightest for studies of systems in
which the neutron star should have accreted substantial amounts of
matter.  While for individual systems the evidence that neutron stars
can have high masses is weak, the ensemble appears to show that masses
around $1.6\,M_\odot$ are possible.  For the radius determination, most
attempts have focussed on neutron stars in low-mass X-ray binaries in
which accretion has temporarily shut down.  These neutron stars are
easiest to model, since they should have pure Hydrogen atmospheres and
low magnetic fields.  To obtain accurate radii, however, requires
precise distances and very high quality data.
}

\section{Trying to constrain the equation of state}

The physics of matter at ultra-high density is not only of interest on
its own accord, but also because of its astronomical implications: to
understand the core collapse of massive stars, the supernova
phenomenon, and the existence and properties of neutron stars,
knowledge of the physics, as summarised in the equation of state
(EOS), is required.  As is clear from other contributions to these
proceedings, quantum-chromodynamics calculations are not yet developed
well enough to determine the densities at which, e.g., meson
condensation and the transition between the hadron and quark-gluon
phases occur.  At densities slightly higher than nuclear and at high
temperatures, the model predictions can be compared with the results
of heavy-nuclei collision experiments.  The substantial progress on
this front is discussed elsewhere in these proceedings.  For higher
densities and low temperatures, however, no terrestrial experiments
seem possible; the models can be compared only with neutron-star
parameters.  Recent reviews of our knowledge of the EOS, and the use
of neutron stars for constraining it, are given by Heiselberg \&
Pandharipande\cite{hp00}, and Lattimer \& Prakash.\cite{lp00,lp01}

Here, I focus on two possible ways to constrain the EOS: aiming to find
the highest observed neutron-star mass, and to measure precise radii.
The first part is an update of reviews I have given
earlier.\cite{vker01a,vker01b}

\section{Maximum mass}

Observationally, after spin periods, masses are perhaps the easiest
bulk properties to determine.  Their possible interest for
constraining the EOS is that for any given EOS, there is a maximum
mass a neutron star can have; beyond this, it would collapse to a
black hole.  For instance, for EOS with a phase transition at high
densities, such as Kaon condensation,\cite{bb94} only neutron stars
with mass $<\!1.5\,M_\odot$ could exist.  This EOS would be excluded
if a neutron star with a mass above this maximum were known to exist.
Below, I first describe the constraints given by the very accurate
masses derived from relativistic binary neutron stars, and next the
less precise masses from X-ray binaries.  I then discuss whether the
narrow mass range implied by the most accurate masses implies a
constraint on the EOS, or rather reflects the astrophysical processes
by which neutron stars form.  I conclude by briefly describing the
situation for neutron stars for which on astronomical grounds one
might expect that they have accreted a substantial amount, and hence
have become more massive.

\subsection{Relativistic neutron-star binaries}

The most accurate mass determinations have come from radio timing
studies of pulsars (see Thorsett \& Chakrabarty\cite{tc99} for an
excellent review).  The best among these are for pulsars that are in
eccentric, short-period orbits with other neutron stars, in which
several non-Keplerian effects on the orbit can be observed: the
advance of periastron, the combined effect of variations in the
second-order Doppler shift and gravitational redshift, the shape and
amplitude of the Shapiro delay curve shown by the pulse arrival times
as the pulsar passes behind its companion, and the decay of the orbit
due to the emission of gravitational waves.  The most famous of the
double neutron-star binaries is the Hulse-Taylor pulsar, PSR B1913+16,
for which $M_{\rm{}PSR}=1.4411\pm0.0007\,M_\odot$ and
$M_{\rm{}comp}=1.3874\pm0.0007\,M_\odot$ was derived.\cite{tw89,t92}
Almost as accurate masses have been inferred for PSR B1534+12, for
which the pulsar and its companion were found to have very similar
mass:\cite{sttw02} $M_{\rm PSR}=1.3332\pm0.0010\,M_\odot$ and $M_{\rm
comp}= 1.3452\pm0.0010\,M_\odot$.

To these two best-known systems, recently two further interesting
binaries have been added.  The first is PSR J1141$-$6545, for which
$M_{\rm{}PSR}=1.30\pm0.02\,M_\odot$ and
$M_{\rm{}comp}=0.99\pm0.02\,M_\odot$ was measured;\cite{bokh03} here,
the companion is most likely a massive white dwarf.\footnote{Given that
this system has an eccentric orbit, it seems certain that the white
dwarf was formed before the supernova explosion that left the neutron
star (and made the orbit eccentric).  Likely, what happened is that
both stars in the binary originally had masses too low to form a
neutron star, but that as the originally more massive star evolved, a
phase of mass transfer ensued in which the originally less massive
star received sufficient material to push its mass over the limit
required for neutron-star formation (for references, see citations in
Bailes et al.\cite{bokh03}).}  Very recently (after the conference),
the discovery of a neutron-star binary was announced in which both
components are radio pulsars.\cite{lyn+04} From this `double-lined'
radio pulsar binary, masses $M_{\rm PSR,1}=1.337\pm0.005\,M_\odot$ and
$M_{\rm PSR,2}=1.250\pm0.005\,M_\odot$ were derived.

From the above, one sees that all these well-determined masses are in
a relatively narrow range, between 1.25 and 1.44$\,M_\odot$.

\subsection{The high-mass X-ray binary Vela X-1}

Neutron-star masses can also be determined for some binaries
containing an accreting X-ray pulsar, from the amplitudes of the X-ray
pulse delay and optical radial-velocity curves in combination with
constraints on the inclination (the latter usually from the duration
of the X-ray eclipse, if present).  This method has been applied to
about half a dozen systems.\cite{jr84,vkvpz95} The masses are
generally not very precise, but are consistent with
$\sim\!1.4\,M_\odot$ in all but one case.

The one exception is the X-ray pulsar Vela X-1, which is in a 9-day
orbit with the B0.5\,Ib supergiant HD~77581.  For this system, a
rather higher mass of around $1.8\,M_\odot$ has consistently been
found ever since the first detailed study in the late
seventies.\cite{vpzt+77,vkvpz+95} A problem with this system, however,
is that the measured radial-velocity orbit, on which the mass
determination relies, shows strong deviations from a pure Keplerian
radial-velocity curve.  These deviations are correlated within one
night, but not over longer periods.  A possible cause could be that
the varying tidal force exerted by the neutron star in its eccentric
orbit excites high-order pulsation modes in the optical star which
interfere constructively for short time intervals.

\begin{figure}[t!]
\centerline{\includegraphics[width=0.9\hsize]{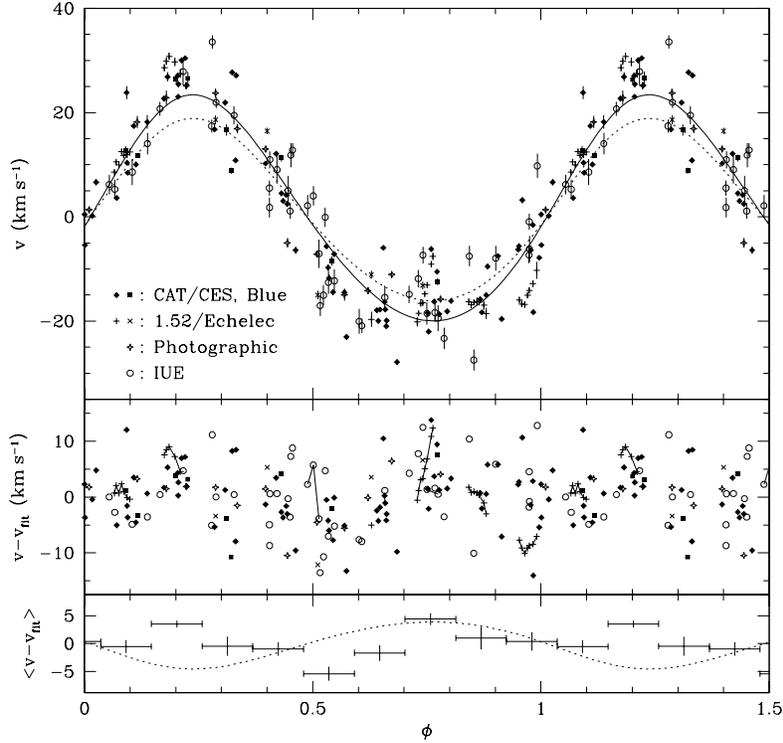}}
\caption[]{Radial-velocity measurements for HD\,77581, the optical
counterpart to Vela X-1.  Overdrawn is the Keplerian curve that best
fits the nightly averages of the data (solid line; $K_{\rm
opt}=21.7\pm1.6{\rm\,km\,s^{-1}}$), as well as the curve expected if
the neutron star has a mass of $1.4\,M_{\odot}$ (dotted line; $K_{\rm
opt}=17.5{\rm\,km\,s^{-1}}$).  The residuals to the best-fit are shown
in the middle panel.  For clarity, the error bars have been omitted.
Points taken within one night are connected with lines.  In the bottom
panel, the residuals averaged in 9 phase bins are shown.  The
horizontal error bars indicate the size of the phase bins, and the
vertical ones the error in the mean.  The dotted line indicates the
residuals expected for a $1.4\,M_{\odot}$ neutron star.}
\label{fig:vela}
\end{figure}

The velocity excursions appeared not to depend on orbital phase, and
we hoped that, with sufficient observations, it would be possible to
average them out.  For that purpose, we obtained about 150 spectra,
taken in as many nights and covering more than 20 orbits, of the
optical counterpart.\cite{bkvk+01} Unfortunately, we found that the
average velocity curve does show systematic effects with orbital phase
(see Fig.~\ref{fig:vela}), which dominate our final uncertainty.
While our best estimate still gives a high mass, of $1.86\,M_\odot$,
the $2\sigma$ uncertainty of $0.33\,M_\odot$ does not allow us to
exclude soft equations of state conclusively.

A different approach would be to try to obtain very dense coverage of
the orbit, in the hope that one could model and remove the excursions.
This approach was tried too,\cite{qna+03} and for the two weeks
covered by the observations, it was found to be possible to model the
velocity excursions as a fairly coherent, 2.18-d oscillation (in
contrast to what was the case in earlier observations\cite{vkvpz+95}).
After removal, however, the systematic orbital-phase dependent effects
that plagued our determination became apparent here too, and the final
mass has roughly the same value, but also roughly the same uncertainty
as the one described above.

\subsection{Physical or astrophysical implications?}

While we cannot draw a firm conclusion about the mass of Vela X-1, it
is worth wondering how it could be the only neutron star with a mass
so different from all others.  I would argue one should be careful in
taking the narrow mass range around $1.4\,M_\odot$ as evidence for an
upper mass limit set by the EOS.\cite{bkvk+01} After all, for all EOS,
neutron stars substantially {\em less} massive than $1.4\,M_\odot$ can
exist, yet none are known.  Could it be that the narrow range in mass
simply reflects the formation mechanism, i.e., the physics of
supernova explosions and the evolution of stars massive enough to
reach core collapse?  There certainly is precedent: white dwarfs are
formed with masses mostly within a very narrow range
around~$0.6~M_\odot$, well below their maximum (Chandrasekhar) mass.

Interestingly, from evolutionary calculations,\cite{tww96} it is
expected that single stars produce neutron stars with a bimodal mass
distribution, with peaks at 1.27 and $1.76~M_{\odot}$.  For stars in
binaries, only a single peak at $\sim\!1.3\,M_\odot$ was found, but it
is not clear whether this result will hold (S.\ Woosley, 2000, pers.\
communication).  If not, could it be that the progenitor of Vela X-1
was a star that managed to produce a massive neutron star?  If so, one
may still wonder why no massive radio pulsars or pulsar companions
have been found.  This may be a selection effect:\cite{bkvk+01} all
neutron stars with accurate masses are in binary neutron stars systems
in close orbits, whose formation requires a common-envelope stage.
During this stage, a merger can only be avoided if the initial orbit
was very wide.  Stars massive enough to form a massive neutron star,
however, likely do not evolve through a red-giant phase, and a
common-envelope phase would occur only for rather close orbits, for
which the binary would merge.

\subsection{Trying for bias}

In considering the mass measurements discussed above, it should be
noticed that for all neutron stars with good masses, it is expected
that they accreted only little mass after their formation.  Only
neutron stars in low-mass X-ray binaries and radio pulsars with
low-mass white dwarf companions are expected to have accreted
substantial amounts of material.  It may thus be worthwhile to try to
bias oneself to more massive neutron stars by studying these.

For low-mass X-ray binaries, higher masses, of $\sim\!2\,M_\odot$,
have indeed been suggested; e.g., from dynamical measurements and
lightcurve fitting for Cyg~X-2,\cite{ok99} and from inferences based
on quasi-periodic oscillations.\cite{zss97} These estimates, however,
rely to greater or lesser extent on unproven assumptions.

The radio pulsars with white dwarf companions provide cleaner systems,
for a number of reasons.  First, most of the radio pulsars in these
binaries spin very rapidly and stably, making them ideally suited for
precision timing.  Second, both components can safely be approximated
as point masses for dynamical purposes, so that deviations from
Keplerian motion can be readily interpreted.  Third, by taking optical
spectra of the white dwarf, it is possible to determine its
radial-velocity curve, and thus obtain the mass ratio.  Finally, from
a model-atmosphere analysis of optical spectra, one can infer the
surface gravity and, using the white-dwarf mass-radius relation, the
mass.  Combined with a mass ratio, this yields the pulsar mass.

\begin{figure}
\centerline{\includegraphics[width=0.9\hsize]{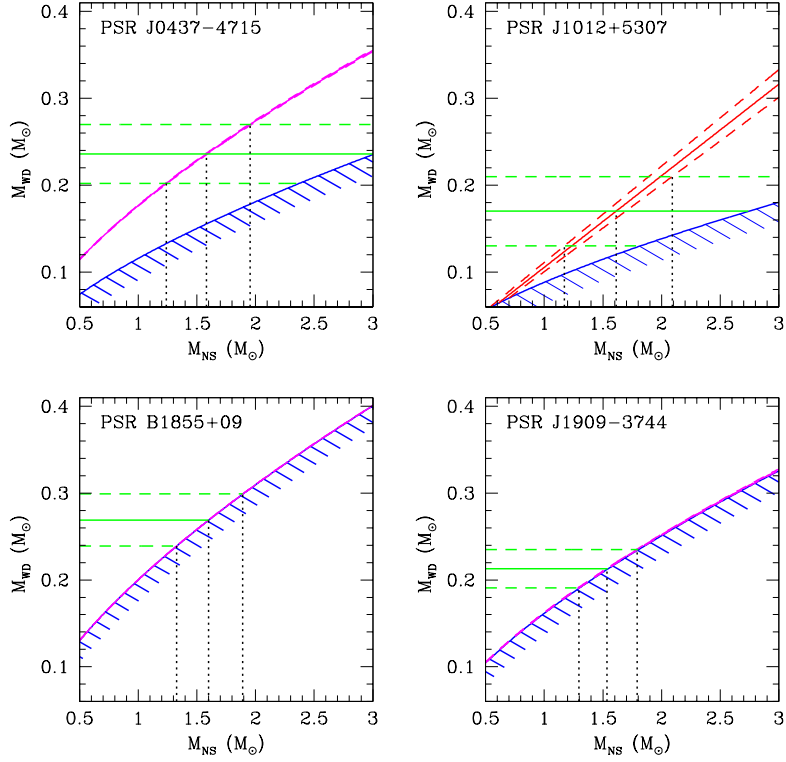}}
\caption[]{Constraints on the masses of four radio pulsars with
white-dwarf companions.\cite{vsbb+01,vkbk96,cgk98,ktr94,nss03,jbvk+03}
In all panels, the solid line with hashing on the lower right reflects
the limit set by the pulsar mass function.  The horizontal lines
reflect the limits on the companion mass, with the 2-$\sigma$
uncertainties.  For PSR J1012+5307, the mass is inferred from a fit to
the optical spectrum, while for the other three sources it is measured
from Shapiro delay.  The Shapiro delay also implies inclinations very
close to $90^\circ$ (edge-on) for PSR B1855+09 and PSR
J1909$-$3744. For PSR J0437$-$4715, the curved line is the constraint
on the inclination, inferred from the change in aspect with which we
view the system due to its proper motion.  For PSR J1012+5307, the
diagonal straight lines show the mass ratio and its 2-$\sigma$
uncertainties, inferred from the radial-velocity amplitudes of the two
components.  The vertical lines show the resulting 2-$\sigma$
constraints on the pulsar masses.\label{fig:fourpsr}} \end{figure}

In many of these systems, one expects the neutron star to have
accreted a substantial amount of matter.  This is because many of the
white dwarf companions have masses of only $\sim\!0.2\,M_\odot$.  In
order for their progenitors to have evolved off the main sequence in a
Hubble time, the initial masses much have been at least
$0.8\,M_\odot$.  Hence, at least $0.6\,M_\odot$ was lost.  From
evolutionary considerations, one would expect much of this to have
been accreted to the neutron star, as mass transfer should have been
relatively slow and stable.

Despite the above expectations, initial results showed no evidence for
such high masses.\cite{tc99} Current results for four systems are
shown in Fig.~\ref{fig:fourpsr}.  One sees that for none of these, the
mass measurements exclude $1.4\,M_\odot$ at high significance, but for
all the best values are above it. This is true for a number of other
systems as well.\cite{nss03}

While the masses appear to be above the narrow range inferred from the
relativistic binary pulsars, suggesting that some accretion has
happened, the masses are not as high as the expected
$\sim\!2\,M_\odot$.  Could this reflect a hard limit set by the EOS?
Unfortunately, there may also be a more mundane, astronomical
explanation.  The pulsars in these binaries have magnetic fields that
are much weaker than those inferred for regular radio pulsars.  On
empirical grounds, it is believed this reduction is related to being
in a binary, with the magnetic field being reduced somehow by the
accretion (for a review, see Phinney \& Kulkarni\cite{pk94}).  If this
is indeed the case, above some amount of accreted matter, the magnetic
field may be reduced so far that no pulsar will be seen.  This would
thus lead one to find a upper limit to the mass of pulsars that
accreted matter.

Even though the above may mean a low apparent upper limit to the mass
distribution may not be very meaningful, the masses do appear to be
higher than in other systems.  At present, therefore, the radio
pulsars with white dwarf companions still seem to be the most
promising systems for obtaining precise and reliable masses that
provide strong constraints on the EOS.  Further radio timing
observations, as well as optical spectroscopy, are underway.

\section{Atmospheric modelling}

Spectroscopic analysis of emission arising in the photosphere of a
neutron star offers, in principle, the possibility of much stronger
constraints.  From models, the temperature and angular diameter could
be inferred, which, combined with a distance, gives the radius.  If
absorption lines are present, also the gravitational redshift
($\Rightarrow M/R$) and pressure broadening ($\Rightarrow M/R^2$) can
be measured.  An accurate measurement of even just one of these may be
useful, given the narrow range of observed neutron-star masses for
neutron stars that did not accrete much after their formation.

In practice, most neutron stars are unsuitable: many have spectra
contaminated or even dominated by poorly understood magnetospheric or
accretion processes.  Even without those, the strong magnetic fields
present in many neutron stars imply that the microphysics (energy
levels, radiative transfer, etc.) is complicated and that the
temperature will in general not be uniform.  Finally, most have
atmospheres almost certainly composed of pure Hydrogen, since, if any
is present, gravitational settling ensures that it will quickly float
to the top.  

\subsection{Quiescent emission from low-mass X-ray binaries}

The only neutron stars for which no pulsations are seen, and which
therefore may not have significant magnetic fields, are those that
reside in low-mass X-ray binaries.  For some of these, the accretion
is episodic, and when accretion turns off, one may be able to observe
simple thermal emission from the surface.\cite{bbr98}

Given that the accreted matter contained Hydrogen, which will float to
the top, the atmosphere should be composed of pure Hydrogen.  This
makes the modelling relatively easy, even if it also implies likely no
lines will be seen (small features may be visible due to continuing
low-level accretion).  Thus, the best one can hope for is to determine
effective temperatures and angular diameters.  If one has a distance,
e.g., because the source is in a globular cluster, this will yield the
radius.  This may even be quite precise, as distances to globular
clusters are becoming better and better.

A possible problem lies in the interpretation: one measures the radius
as seen at infinity, $R_\infty = R/\sqrt{1-2GM/Rc^2}$, which,
unfortunately, depends more strongly on the precise value of the mass
than the radius itself.  Since accretion has happened, one does not
know the mass a priori.  Nevertheless, significant constraints are
possible.  For instance, for many soft EOS, one finds
$R_\infty<14\,$km for neutron stars with any mass between 0.5 and the
upper limit for those EOS, $\sim\!1.6\,M_\odot$.  Similarly, for the
stiffer EOS, one finds $R_\infty>15\,$km for any mass above
$1.35\,M_\odot$.

Attempts to measure radii for neutron stars in low-mass X-ray binaries
in quiescence have been made by a number of authors.  Most work has
been done for binaries in the
field.\cite{bbr98,rbb+99,rbb+01,rbb+02b,wgvdkm02} For these, one will
need to obtain parallaxes to obtain meaningful radii.  However, the
observations also showed that the interpretation was not always as
straightforward as hoped.  First, for some sources, interstellar
absorption cuts off a large part of the spectrum, making it difficult
to measure the temperature accurately.  Second, in the best studied
source, Aql X-1, a non-thermal component was found, and even the
thermal component was found to be variable on a relatively short
timescale of months.\cite{rbb+02a} These complications shed some doubt
on whether the observations are interpreted correctly.  It also gives
some hope, however: the non-thermal emission is likely due to residual
accretion, and it may be possible to see the signature of that in
lines from heavier elements.

More recently, the focus has shifted to sources in globular clusters,
where the distances are known to fair precision.  The best results
have come from XMM, because of its large collecting
area.\cite{gbw03a,gbw03b} For the two sources studied in most detail,
the uncertainty in the radius now seems dominated by the uncertainty
in the distance; the uncertainty due to just the X-ray fitting appears
to be below 0.5\,km.

\subsection{X-ray bursts}

The X-ray emission of low-mass X-ray binaries is dominated by emission
from the neutron-star surface not only in quiescence, when accretion
is absent, but also during so-called X-ray bursts.  These X-ray bursts
occur when the layer of freshly accreted matter becomes unstable to
nuclear burning (usually of Helium; Hydrogen is burnt at least
partially as matter accretes; for a review, see Bildsten\cite{bil00}).
The bursts typically last a few seconds, and occur every few hours.

In a very long, 335\,ks {\it XMM} observation of the X-ray burst
source EXO~0748$-$676, taken for calibration purposes, 28 X-ray bursts
were observed.  By analysing the summed spectra obtained during these
bursts (which lasted a cumulative 3.2\,ks), possible small absorption
features were found, which could be identified with the $n=2$ to 3
transition of Hydrogen- and Helium-like Iron,\cite{cpm02} for a
gravitational redshift $z=0.35$.  It is not clear where the Iron
originates.  One might think it results from the nuclear fusion, and
is brought to the surface, but an alternative explanation is that the
metals observed are brought to the photosphere by continuing
accretion.\cite{bcp03}

Unfortunately, while the results are extremely intriguing, they are
very difficult to confirm, since observationally it is hard to
get much higher quality data.

\section{Conclusions and prospects}

The pessimistic conclusion from the above would be that, since their
discovery, neutron stars have not much advanced our understanding of
the physics at extreme densities.  Viewing the situation more
positively, though, one sees that the mass determinations are now
getting accurate enough to be interesting, especially for the radio
pulsars with white dwarf companions.  The precision will increase with
further timing of the pulsars, helped by optical studies of
the white dwarfs.

Furthermore, new avenues are explored in which the thermal emission is
modelled and used to derive angular diameters and gravitational
redshifts.  The study of low-mass X-ray binaries in quiescence seems
particularly promising, particularly once the distances settle down.
For globular clusters, this is in progress, while for individual
systems in the field it will come in the somewhat longer run, using
direct parallax determinations with NASA's Space Interferometry
Mission or ESA's GAIA, both expected to be launched in about ten
years.

Finally, not mentioned in my talk or this write-up, there are a number
of nearby neutron stars with what appear to be purely thermal spectra.
Recently, for a number of these, absorption features have been
detected, which are likely due to Hydrogen atmospheres in extremely
strong magnetic fields, $10^{13}$ to $10^{14}\,$G.\cite{vkkd+04} In
such fields, absorption might be due to proton cyclotron or neutral
Hydrogen; with more than a single line, one may be able to solve for
the field strength and the gravitational redshift.  Several are close
enough for parallax determinations; combined with angular diameters
from atmospheric modelling, this might yield good radii.

Overall, the outlook seems fairly bright.

\def\aap,#1,{{\it Astron.\ Astroph.{}} {\bf #1},}
\def\aaps,#1,{{\it Astron.\ Astroph.\ Supp.{}} {\bf #1},}
\def\apj,#1,{{\it Astroph.\ J.{}} {\bf #1},}
\let\apjl=\apj
\def\araa,#1,{{\it Ann.\ Rev.\ Astron.\ Astroph.{}} {\bf #1},}
\def\mnras,#1,{{\it Mon.\ Not.\ R.\ Astron.\ Soc.{}} {\bf #1},}
\def\nat,#1,{{\it Nature} {\bf #1},}

\end{document}